\documentclass[letterpaper, 10 pt, conference]{ieeeconf}

\IEEEoverridecommandlockouts                              %
\usepackage{amsmath,amssymb,amsfonts}
\usepackage{algorithmic}
\usepackage{algorithm}
\usepackage{xcolor}
\usepackage{graphicx}
\usepackage{wrapfig}
\usepackage[hidelinks]{hyperref}
\usepackage[capitalize, nameinlink]{cleveref}

\title{\LARGE \bf
    Graph Neural Network-based Multi-agent Reinforcement Learning for Resilient Distributed Coordination of Multi-Robot Systems
}

\author{Anthony Goeckner$^1$, Yueyuan Sui$^1$, Nicolas Martinet$^{1,2}$, Xinliang Li$^1$, and Qi Zhu$^1$
    \thanks{$^{1}$Department of Electrical and Computer Engineering,
            Northwestern University, Evanston, Illinois, USA. Corresponding author: Anthony Goeckner ({\tt\small anthony.goeckner@northwestern.edu}).}%
    \thanks{$^{2}$CNRS, Inria, Rennes, France. 
            }
}

\begin{document}
\maketitle
\thispagestyle{empty}
\pagestyle{empty}

\begin{abstract}
Existing multi-agent coordination techniques are often fragile and vulnerable to anomalies such as agent attrition and communication disturbances, which are quite common in the real-world deployment of systems like field robotics. To better prepare these systems for the real world, we present a graph neural network (GNN)-based multi-agent reinforcement learning (MARL) method for resilient distributed coordination of a multi-robot system. Our method, Multi-Agent Graph Embedding-based Coordination (MAGEC), is trained using multi-agent proximal policy optimization (PPO) and enables distributed coordination around global objectives under agent attrition, partial observability, and limited or disturbed communications. We use a multi-robot patrolling scenario to demonstrate our MAGEC method in a ROS~2-based simulator and then compare its performance with prior coordination approaches. Results demonstrate that MAGEC outperforms existing methods in several experiments involving agent attrition and communication disturbance, and provides competitive results in scenarios without such anomalies.
\end{abstract}

\section{Introduction}
Multi-agent systems surround us every day: vehicles on the road, airplanes in the sky, and soon, robots in multitudes. For truly useful autonomy in these systems, individual agents must coordinate their actions with those around them. However, current multi-agent coordination techniques are often fragile, susceptible to both drastic changes such as agent attrition or to the slightest disturbances such as poor communications, especially given the inherent partial observability of our physical world. In this paper, we present new work towards a novel graph-based reinforcement learning approach to multi-agent coordination that is robust to such disturbances.

A multi-robot system that depends on inter-agent coordination cannot be fielded in environments which are particularly disturbance-prone or adversarial without robustness to those disturbances. This problem was highlighted extensively in the DARPA OFFSET program, which fielded nearly two hundred autonomous ground and air vehicles in urban combat scenarios \cite{taranta_warfighting_2023}. Experimentation during the OFFSET program revealed significant coordination deficiencies in the presence of frequent agent attrition and a difficult communication environment. Most existing works do not address these disturbances effectively or at all.

Many real-world environments (and associated problems) can be naturally represented by graphs. For example, a road network (route planning), dense forest (swarm motion planning), or warehouse (task assignment) are graph environments at heart. These environments are then well suited for modeling and analysis using Graph Neural Networks (GNNs), which aggregate information from throughout the graph using a ``message-passing'' process to learn an ``embedding'', or representation, of the graph. This allows GNNs to capture the attributes of individual entities in the graph and the complex relationships between them.

The combination of GNNs and multi-agent reinforcement learning (MARL) is therefore a natural fit for solving many multi-robot challenges.
Until now though, GNNs have not been heavily used in MARL and multi-robot coordination tasks.


We aim to change this using MAGEC, the Multi-Agent Graph Embedding-based Coordination algorithm. It is a MARL framework based on multi-agent proximal policy optimization (MAPPO) and inductive GNNs. Our approach is capable of multi-agent coordination in the face of agent attrition, poor or nonexistent communication, and partial observability of the environment. It is applicable to the general class of problems in which agents must traverse a graph-based environment. In this paper, we apply MAGEC to the multi-robot patrolling scenario, which we evaluate using our custom multi-robot simulation environment. Our evaluation includes comparison with benchmark algorithms for the patrolling scenario and provides analysis of our various design decisions. \textbf{Our contributions are as follows:}

\begin{itemize}
    \item MAGEC, a robust GNN-based MARL method for multi-agent coordination in the face of disturbances, delayed rewards, and sparse actions.
    
    \item An inductive graph embedding method that accounts for both node and edge features.

    \item A novel ``neighbor scoring'' GNN which enables navigation in a graph-based environment.

    \item Training and simulation tools for MARL-based multi-robot patrolling.
\end{itemize}





\section{Related Works}
\label{sec:related}

\subsection{Graph Neural Networks}
Many multi-robot coordination problems, including ours, can be easily represented by a graph. However, traditional neural networks accept only fixed-size inputs such as bitmap observations. In the case of a graph, the input size may vary based on the graph structure. To handle this, the so-called ``message-passing'' paradigm was developed first in \cite{gilmer_neural_2017} and then in many later works such as \cite{yao_improving_2023} to learn graph embeddings.

However, many of these prior works represent graphs in a transductive manner and do not generalize well to graphs different from those on which the embedding was originally trained. To overcome this challenge, Hamilton et al. devise an \textit{inductive} embedding method called ``GraphSAGE'' \cite{hamilton_inductive_2018}. This method was further improved upon by Xu et al. in \cite{xu_representation_2018} with the addition of skip connections between GraphSAGE convolutional layers.

Zhou et al. proposed a GNN framework \cite{zhou_co-embedding_2023} that integrates node and edge features to boost long-distance information transmission and enhance node embeddings.
Our independently developed research also adopts combined node and edge embedding, though our method is inductive and may be applied to graphs and nodes unseen during training.

\subsection{Multi-Agent RL using Graph Neural Networks}
Compared to other methods, the use of GNNs in MARL has been minimal. In \cite{zhang_neural_2023}\cite{zhang_gcbf_2024}, a single-layer GNN is used to interpret LIDAR observations. In \cite{naderializadeh_graph_2021, ding_multiagent_2023}, a GNN is used during the training stage of a MARL algorithm to perform state-action value factorization. Hu et al. use a GNN for learned multi-robot coordination, though they only consider the agents and not the environment as part of the graph \cite{hu_graph_2023}.

\subsection{Algorithms for Patrolling}
A series of studies have progressively enhanced understanding of and methodology for the multi-robot patrolling problem. Portugal and Rocha proposed two Bayesian-based strategies, namely the \textit{Greedy Bayesian Strategy} (GBS) and the \textit{State Exchange Bayesian Strategy} (SEBS) \cite{portugal_distributed_2013}. GBS emphasized immediate gains by making locally optimal choices and SEBS extended this by integrating inter-robot communication attrition robustness. Portugal and Rocha later introduced the \textit{Concurrent Bayesian Learning Strategy} (CBLS) \cite{portugal_cooperative_2016}, which utilized reward-based learning. Farinelli et al. developed two solutions for patrolling in uncertain environments: DTA-Greedy and the more intricate DTAP, an auction-based task allocation mechanism \cite{farinelli_distributed_2017}. Wiandt et al. presented a self-organizing algorithm that could autonomously partition a graph for agent patrolling \cite{wiandt_self-organized_2017}.
Kobayashi et al. proposed the Local Reactive (LR) algorithm \cite{kobayashi_multi-robot_2023}, which focuses on patrolling while maintaining base station situation awareness.
More recently, ElGibreen and Youcef-Toumi introduced an online DTA method for uncertain environments, emphasizing heterogeneous agent capabilities \cite{elgibreen_dynamic_2019}. The \textit{Adaptive Heuristic-based Patrolling Algorithm} (AHPA) was introduced in \cite{goeckner_attrition-aware_2024} and was shown to be robust to agent attrition while using minimal communication resources.

\subsection{Algorithms for MARL Patrolling}
More recently, MARL has been used for multi-robot patrolling. Guo et al. use graph attention networks with MARL \cite{guo_balancing_2023}. However, their method does not appear to generalize to different environments without retraining.
Tong et al. present a MARL approach \cite{tong_energy-aware_2023} for patrolling, though it only uses simple grid environments with single-step actions and immediate reward.



\section{Problem Definition}
\label{sec:formulation}
We start by providing the mathematical formulation for our multi-agent system and then introduce the motivating scenario, multi-robot patrolling.

In the following, we use subscript notation to indicate a value pertaining to an agent, and superscript notation to indicate an agent's belief about some value. For example, $s_i^j$ is agent $j$'s belief about the state of agent $i$, and $s_i$ is the true state of agent $i$. When necessary, we denote a value at a specific time $t$.

\subsection{Multi-Agent System Definition}
Using the notation of \cite{boutilier_planning_1996}, we formulate the multi-agent system as a decentralized partially-observable Markov decision process (Dec-POMDP) described by the tuple $\langle \mathcal{A}, S, O, A, T, R, \gamma \rangle$.
The set $\mathcal{A}$ is a finite set of agents. As shorthand, we let $n = |\mathcal{A}|$.

The environment is modeled as a graph $G = \{V, E\}$ in Euclidean space $\mathbb{R}^2$, where $V$ is the set of vertices and $E$ is the set of edges. As shorthand, we let $m = |V|$ indicate the number of vertices.

The set $S$ is a finite set of possible system states and
$$s = \{s_i\}_{i \in \mathcal{A}} \cup \{s_{v}\}_{v \in V} \enskip \forall s \in S$$
where $s_i$ is the state of agent $i$ and $s_v$ the state of node $v$. Due to partial observability and communication disturbances, the true state $s$ of the system may not be known with certainty. Rather, each agent only knows with certainty a subset of the system state, $s^i$.

The set $O$ is the joint observation space, where $O = \{O_i \}_{i \in \mathcal{A}}$. The observation space $O_i$ of each agent is described as the state of all vertices and agents within some radius $r$ of the agent $i$.

Information is shared between agents by observation and explicit communication. Agent $i$ may observe the state $s_j$ of other agents and vertices within radius $r$:
$$o_i = \{s_j | j \in V \cup \mathcal{A} : dist(i, j) \leq r\}$$

Agents may only travel between nodes which are joined by an edge in $E$, and travel is considered an action in $A_i$, the set of available actions for agent $i$. The action space is defined as $A_i = \{0 ,\dots ,\Delta(G)-1\}$, where $\Delta(G)$ is the maximum degree of graph $G$. We also define $A = \{A_i \}_{i \in \mathcal{A}}$ as the joint action space over all agents.

A transition function $T: S \times A \times S \to [0, 1]$ describes the probability of transition from one system state to another given some joint action.
Finally, the function $R: S \to \mathbb{R}^n$ describes the reward to each agent for a particular system state, and $\gamma$ is a discount factor on the reward.
We assume that $T$ and $R$ are defined as part of the environment and are unknown a priori.



The optimization objective of the agent is to find a policy $\pi_i(s^i)$ which provides the agent's best action given the estimated system state $s^i$. We assume that spaces $S,O,A$ are known in advance by all agents and that the policy $\pi_i$ may be shared such that it is the same on all agents: $\pi^i=\pi^j \enskip\forall i,j \in \mathcal{A}$. Then, we may train a policy using centralized training and decentralized execution (CTDE).


\subsection{The Patrolling Problem}
\label{sec:patrolling}


Multi-Agent Patrolling is a common benchmark problem for evaluation of coordination algorithms such as task allocation \cite{portugal_distributed_2013}. We use the patrolling formulation of \cite{goeckner_attrition-aware_2024} and adapt it to this problem. In the patrolling problem, agents must repeatedly visit a set of \textit{observation points} or nodes, attempting to minimize one of several metrics such as $\bar{\zeta} = \frac{1}{m} \sum_{v \in V} \zeta(v)$, the average idleness time or $\zeta_\text{max} = \max_{v \in V} \zeta(v)$, the worst node idleness time.

Unlike in some other patrolling methods, such as \cite{portugal_distributed_2013}, we assume that the last visitation time of nodes can be observed from the environment (see \cref{sec:observations}).


\section{Design Methodology}
\label{sec:design}
To overcome the serious and realistic limitations imposed by the problem formulation, including partial observability, disturbed communications, and agent attrition, we present our Multi-Agent Graph Embedding-based Coordination (MAGEC) approach. MAGEC uses an actor-critic architecture, with a custom $k$-convolution GNN serving as the actor. Given graph-based inputs, the actor must select an edge of the graph for the agent to next traverse.

Training of MAGEC is performed in an entirely inarcane manner by means of the multi-agent proximal policy optimization (MAPPO) algorithm \cite{yu_surprising_2022}, with some modifications.

\subsection{Overall Architecture}
We must enable distributed coordination while still optimizing towards a global objective. Therefore, we develop a reinforcement learning architecture based on multi-agent proximal policy optimization (MAPPO) \cite{yu_surprising_2022}. We leverage the centralized training and decentralized execution (CTDE) paradigm to enable distributed coordination among the agents. After training is complete, the actor network is used as a policy mapping states to actions, while the critic is discarded. Training uses a shared and omniscient critic $\hat{V}$ for all agents, enabling optimization of agent policies based on the global objective (see \cref{sec:patrolling}). All agents use a shared policy $\pi$, allowing for varying numbers of agents and for adaptation to agent attrition.

\begin{figure}[h]
    \vspace{2mm}
    \centering
    \includegraphics[width=0.45\textwidth]{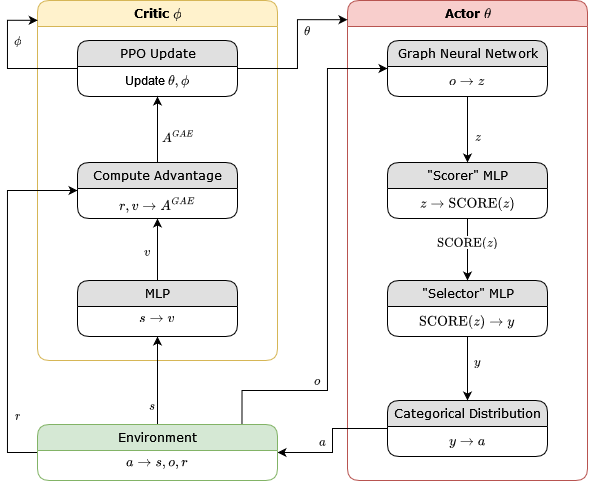}
    \caption{The overall MAGEC training architecture is seen above. Note that the critic is only used during training (CTDE). Please see \cref{fig:gnn} for details of the GNN block.}
    \label{fig:architecture}
\end{figure}

The actor and critic networks have different architectures, reflecting their distinct purposes. The critic network is extraordinarily simple, consisting only of a multi-layer perceptron (MLP). It takes as input the global information necessary to judge overall performance, using feature-engineered information such as the normalized idleness time of all nodes and an adjacency matrix. Its output is simply the value of the current state. This simple architecture enables the critic to provide good estimations of value, but it is far too simple to enable effective coordination for agents.

Therefore, we develop an actor neural network which is far more capable than the critic. The actor is based on the message-passing GNN paradigm, which we describe along with our GNN architecture in \cref{sec:gnn}. First, a graph-based observation (\cref{sec:observations}) is passed into the GNN. Neighbor scoring is then performed on the GNN output using an MLP (\cref{sec:neighbor_scoring}), and the output of neighbor scoring is then passed through another MLP, the ``selector''. The output of this ``selector'' MLP is interpreted as a categorical distribution over the discrete possible actions. Each action represents an edge connected to the agent's current node which the agent should follow (\cref{sec:wayfinding}).

The overall architecture is presented in \cref{fig:architecture}, and proceeding sections describe components in greater detail.

\subsection{Discrete Wayfinding in a Graph}
\label{sec:wayfinding}

\begin{wrapfigure}{L}{0.25\textwidth}
    \centering
    \includegraphics[width=0.25\textwidth]{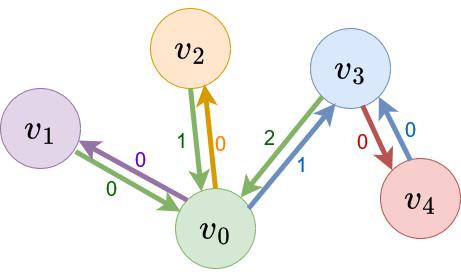}
    \caption{An example of neighbor indexing. Note that $v_3$ is neighbor $2$ of $v_0$, but $v_0$ is neighbor $1$ of $v_3$. Neighbor indexing is enforced by the environment observation mechanism.}
    \label{fig:neighbor_indexing}
\end{wrapfigure}

Let $\mathcal{N}(v)$ be the neighbourhood of node $v$. By definition, this set of neighbors has no particular order. However, to enable generalizable discrete wayfinding, we must enforce an ordering upon these neighbors. As discussed in \cref{sec:observations}, we ensure that neighbors of each node are always presented in the same (undefined) order in every observation. Neighbors are assigned identifiers in the range $0 ,\dots ,|\mathcal{N}(v)-1|$ as shown in \cref{fig:neighbor_indexing}. In order for each node's neighbors to use this same range of identifiers, the undirected graph $G$ must first be converted into a bidirected graph. This process is described in more detail in \cref{sec:observations}.

The agent's policy then outputs an edge index for the neighbor to traverse which corresponds with the desired next node to visit. We use action masking to ensure that the action selected is never greater than the degree of the current node. If an agent is already traversing an edge, we use masking to ensure that the only option is to continue until complete.

\subsection{Graph Neural Network Design}
\label{sec:gnn}
Our GNN is based on the GraphSAGE algorithm \cite{hamilton_inductive_2018}\cite{xu_representation_2018}. However, GraphSAGE does not account for edge attributes, which are critical to wayfinding in a graph. The edge attributes in our graph contain both the weight (length) of the edge and an edge identifier as described in \cref{sec:wayfinding}.

\begin{figure}[h]
    \vspace{2mm}
    \centering
    \includegraphics[width=0.45\textwidth]{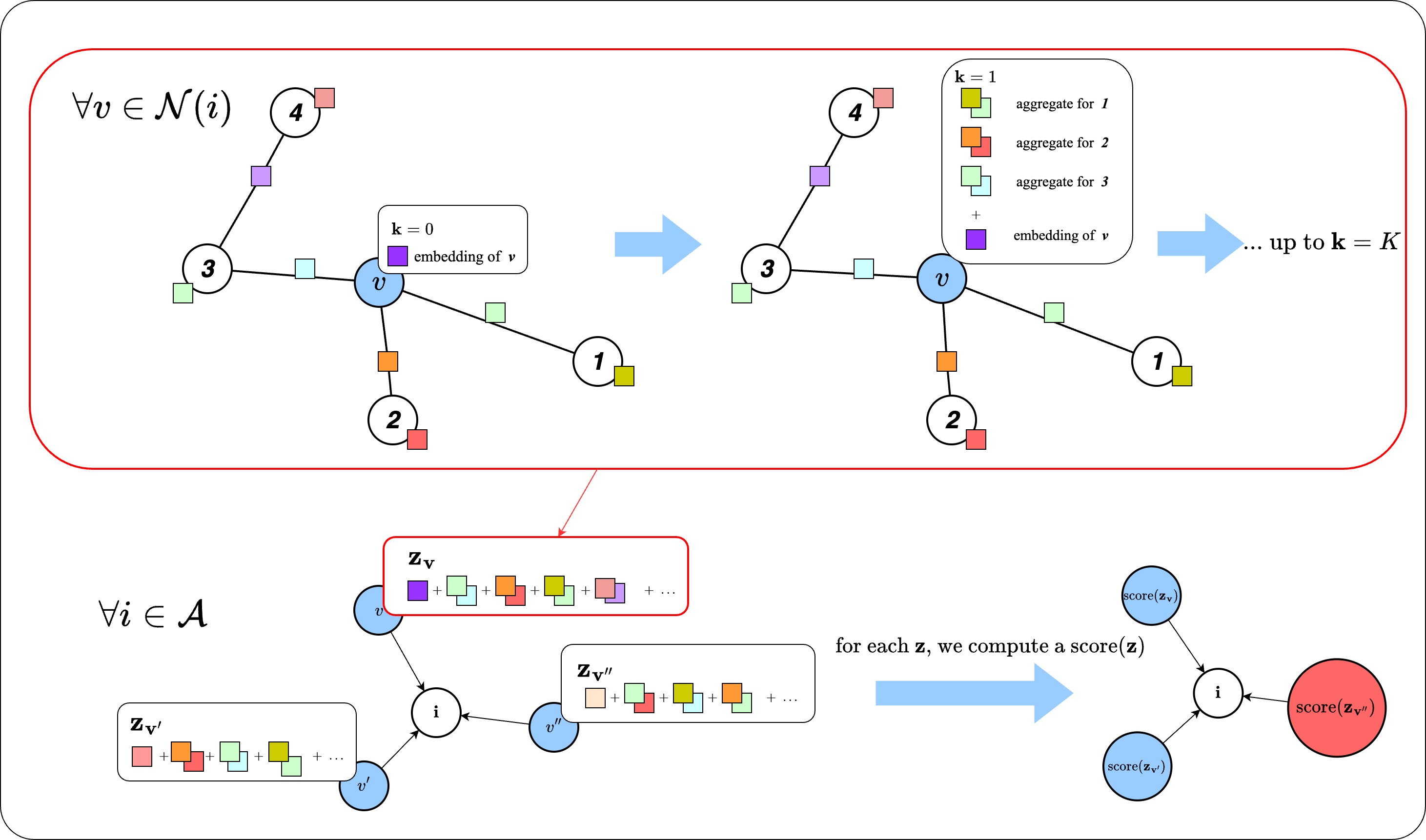}
    \caption{Illustration of GNN computing node embeddings through iterative neighborhood aggregation. A scoring function is applied to the embeddings for decision-making. Note that node and edge features are concatenated during message-passing.}
    \label{fig:gnn}
    \vspace{-5mm}
\end{figure}

To enable consideration of these edge attributes in the convolutional layers, we modify the GraphSAGE embedding generation algorithm as shown in \cref{alg:graphsage_with_edges}. As each ``message'' is passed between nodes, we concatenate the transmitted node features $x_v$ with the features $x_{u,v}$ of the edge that is being traversed to create an augmented feature vector, $\mathring{x}_{u,v}$. Message passing otherwise proceeds as normal in GraphSAGE. This is an efficient and simple method which enables our GNN to consider both node and edge features. Our message-passing framework is shown in \cref{fig:gnn}.

\begin{algorithm}
    \caption{GraphSAGE Embedding Generation (Forward Propagation) with Edge Attributes}
    \label{alg:graphsage_with_edges}
    \begin{algorithmic}[1]
        \STATE \textbf{Input}: Graph $G(V, E)$; input features $\{x_v, \forall v \in V\}$; edge features $\{x_{u,v}, \forall u,v \in V\}$; depth $K$; weight matrices $W_k, \forall k \in \{1, \ldots, K\}$; non-linearity $\sigma$; differentiable aggregator functions $\text{AGGREGATE}_k, \forall k \in \{1, \ldots, K\}$; neighborhood function $\mathcal{N} : v \rightarrow 2^V$
        \STATE \textbf{Output}: Vector representations $z_v$ for all $v \in V$
        \STATE $h_v^0 \gets x_v, \forall v \in V$
        \FOR{$k = 1$ \textbf{to} $K$}
            \FORALL{$v \in V$}
                \STATE $\mathring{x}_{u,v} \gets \text{CONCAT}(h_v^{k-1}, x_{u,v})\ \forall u \in \mathcal{N}(v)$
                \STATE $h_{\mathcal{N}(v)}^k \gets \text{AGGREGATE}_k(\{\mathring{x}_{u,v} \}_{ \forall u \in \mathcal{N}(v)})$
                \STATE $h_v^k \gets \sigma\left(\mathbf{W}_k \cdot \text{CONCAT}(h_v^{k-1}, h_{\mathcal{N}(v)}^k)\right)$
            \ENDFOR
            \STATE $h_v^k \gets \frac{h_v^k}{\|h_v^k\|_2}, \forall v \in V$
        \ENDFOR
        \STATE $z_v \gets h_v^K, \forall v \in V$
    \end{algorithmic}
\end{algorithm}

With each message-passing convolutional layer described in \cref{alg:graphsage_with_edges}, the graph embedding $h_v^k$ gains information about nodes an additional hop from $v$, for all nodes $v \in V$. Therefore, to effectively embed a graph environment in which many hops must be traversed, $k$ must be greater than one. This is in contrast with other GNN-based MARL works, such as \cite{zhang_neural_2023}\cite{zhang_gcbf_2024}, where a value of $k=1$ ensures that only immediate neighbors are represented in the embedding. To balance computational cost and effectiveness of the embedding, we select a value of $k=10$ which enables strong performance in large graphs. As in previous works, we find that skip connections must be included between message-passing layers for efficient training, so we implement ``jumping knowledge'' skip connections as described in \cite{xu_representation_2018}.

In especially large graphs, this fixed value of $k=10$ ensures that the environment is only partial observable, though we find that our method is easily able to overcome such limitation. We also enforce additional partial observability mechanisms during evaluation, described in \cref{sec:observations}.

\subsection{Neighbor Scoring}
\label{sec:neighbor_scoring}
Existing works such as \cite{nayak_scalable_2023} and \cite{zhang_neural_2023} combine graph neural networks with MARL, but they do not attempt to solve the discrete wayfinding problem described in \cref{sec:wayfinding}. To address those challenges involved in selection of an edge to traverse, we devise a strategy which we term, ``neighbor scoring''.

In the forward propagation step of our algorithm, we first create graph embeddings from the perspective of each node using the GNN described in \cref{sec:gnn}. We specify a node of interest, $v$, which represents the agent for which we desire to select an action (see \cref{sec:observations} for more information about agent representations). The graph embedding from the perspective of each neighbor $u \in \mathcal{N}(v)$ is pulled from the output of the GNN and passed through the neighbor-scorer MLP, as seen in \cref{fig:architecture}. This neighbor-scorer MLP provides a single score for each neighboring node. All scores are then passed to a ``selection'' MLP, the output of which forms a categorical distribution over the available actions. We then sample from that categorical distribution to select an action. Borrowing from the notation of \cref{alg:graphsage_with_edges}, the neighbor scoring mechanism may be seen as follows:
$$y_v \sim \text{SELECTION}(\langle \text{SCORE}(z_u) \rangle_{u \in \mathcal{N}(v)})$$

The neighbor scoring mechanism can be seen as a ``chokepoint'' which creates compact latent representations (in our case, a single ``score'' value) of each neighbor's graph embedding. This enables the selection MLP to effectively distinguish between different neighbors' utilities.

We ensure that neighbor scores passed to the selection MLP are always in order of their neighbor index. This enables the selection MLP to effectively learn which embeddings correspond with which neighbor indices, and thus, with which action.

To enable generalization to different numbers of nodes and agents, we train the network with a fixed maximum number of neighbors per node. For nodes with fewer neighbors, we use padding to ensure that the input to the neighbor-scorer MLP remains the same size.

\subsection{Training Environment}
\subsubsection{Observations}
\label{sec:observations}
Agents observe node and edge features of the graph within a limited observation radius. Any observed agents are added to the graph as nodes, with edges indicating the agent's current position relative to other nodes. The graph is then converted to a digraph to enable edge indexing (see \cref{sec:wayfinding}).
Node features for include encoded type (agent or patrol node), normalized node idleness time $\zeta$, and node degree. Edge features include normalized distances and an identifier as described in \cref{sec:wayfinding}.

\subsubsection{Reward}
\label{sec:rewards}
We must optimize for a global objective: the minimization of average idleness time $\bar{\zeta}$. To achieve this, we provide a global ``terminal'' reward $r_{\text{terminal},i}(t) = \frac{t}{\bar{\zeta}} \enskip\forall i \in \mathcal{A}$ to all agents at the end of every episode. We also use a local reward which is awarded to individual agents upon reaching a node $v$. This reward is based on the ratio between the idleness of the visited node $\zeta(v_i)$ and the average idleness of all nodes such that $r_{\text{local}, i}(t) = \frac{\zeta(v_i)}{\bar{\zeta} + \epsilon} \enskip\forall i \in \mathcal{A}$, where $\epsilon$ is a small value which avoids division by $0$. The overall reward function may be written as
$$r_i = \begin{cases}
\alpha \:r_\text{local, i} + \beta \;r_\text{terminal, i} & \text{if $t = \mathcal{T}-1$} \\
\alpha \: r_\text{local, i} & \text{otherwise}
\end{cases}$$

where $\mathcal{T}$ is the preconfigured episode length. This reward $r_i$ is awarded to every agent $i \in \mathcal{A}$ at every step $t = 0, \ldots, \mathcal{T}-1$. In testing, we use $\alpha=1.0$ and $\beta=0.5$.

\subsection{Training Algorithm}
\label{sec:training}
Training of the agents occurs in a centralized fashion following a modified version of the MAPPO algorithm \cite{yu_surprising_2022}.
The rewards provided by our patrolling environment (see \cref{sec:rewards}) are sparse; the agent is only rewarded upon visiting nodes and upon termination of the episode. Further, the patrolling problem described in \cref{sec:formulation} involves \textit{sparse actions} which require multiple time steps to complete. Therefore, after sampling an action $a_i(t)$ from the policy, agent $i$'s next action $a_i(t+\Delta t)$ need not be sampled from the policy until $\Delta t$ steps have passed from $t$. We use this revelation to reduce the number of samples stored in the replay buffer, greatly increasing the speed of training. Unfortunately, due to technical limitations of our implementation, we are unable to skip steps for each agent asynchronously. Rather, we only skip steps synchronously; when one agent needs to sample the policy for a new action, all other agents must also take that step. However, the total number of steps taken in an episode is still greatly reduced thanks to the extremely sparse action nature of our problem.

To account for rewards which are given during skipped steps, we further enhance the algorithm by adding the sum of those rewards to the replay buffer.
$$R_i(s(t), t) = \sum_{k=t}^{t+\Delta t} R_i(s(k), k)$$
The cumulative reward $R_i(s(t), t)$ is then associated with $a_i(t)$ and added to the replay buffer. The rewards for skipped steps $t \dots t + \Delta t$ are not added to the buffer individually.

Similar modifications to the MAPPO algorithm were first made by Yoshitake and Abbeel in \cite{yoshitake_impact_2023}, and we use their adapted version of Generalized Advantage Estimation (GAE) here to great success.
Yoshitake and Abbeel modify GAE equations to account for skipped steps as follows:
$$\delta_i(t) = R_i(s(t), t) + \gamma^{\Delta t} \hat{V}(s(t+1)) - \hat{V}(s(t))$$
$$\text{GAE}_i(t) = \delta_i(t) + (\gamma\lambda)^{\Delta t} \delta_i(t+1) + \cdots + (\gamma\lambda)^{\Delta \mathcal{T}-1} \delta_i(\mathcal{T}-1)$$

As seen above, the discounting factors $\gamma$ are merely raised to the number of steps between samples $\Delta t$. Without skipping, $\Delta t$ is clearly $1$ and the modified GAE and TD equations are then equivalent to the originals.

\begin{algorithm}
    \caption{Training Algorithm}
    \label{alg:training}
    \begin{algorithmic}[1]

        \STATE Initialize Policy $\pi$ to parameters $\theta$
        \STATE Initialize Critic $\hat{V}$ to parameters $\phi$
    
        \FORALL{\text{episodes} $e \in 0 \dots e_\text{max}$}
            \STATE Initialize Rollout buffer $B = \langle s(0), o(0) \rangle$

            \FORALL{$t \in 0 \dots \mathcal{T}$}
                \STATE $\Delta t \gets$ number of steps until next action required
                \STATE $a(t) \gets \pi(o(t))$
                \STATE Take $\Delta t$ steps in environment using action $a(t)$
                \STATE $s(t+\Delta t), o(t+\Delta t) \gets$ values from environment
                \STATE $r \gets \sum_{\tau=t}^{\Delta t} R(s(\tau))$
                \STATE $v \gets \hat{V}(s(t))$
                \STATE $B \mathrel{+}= \langle s(t),o(t),a(t),r,v,\Delta t \rangle$
            \ENDFOR

            \STATE Compute returns $\hat{A}$ using modified GAE over $B$
            \FORALL{mini-batches $b \in B$}
                \STATE Update $\phi$ using mini-batch data $b$.
                \STATE Update $\theta$ using mini-batch data $b$ and critic $\hat{V}$.
            \ENDFOR
        \ENDFOR
    \end{algorithmic}
\end{algorithm}



We find experimentally that use of agent attrition is unnecessary during training and that agents will adapt to attrition during execution nonetheless. This is likely because the problem is formulated as a DEC-POMDP such that the policy $\pi(s)$ has no time dependency and only needs the current state of the system.

\subsection{Execution}
\label{sec:execution}
Use of the trained policies in our patrolling scenario is straightforward and is consistent with typical CTDE practice. Since the critic was only used during training as a surrogate of the value function $V_\pi(s)$, it is not required in the execution phase. The policy trained provides a mapping from state $s(t)$ to appropriate action(s): $\pi(s(t)) \to a(t)$. Therefore, the policy may be straightforwardly applied to agents during execution. Since the policy is shared amongst all agents and operates only on the agent's belief about the global state $s^i(t)$, it may be executed by any number of homogeneous agents regardless of the number of agents used in training.

\subsection{Implementation}
We implement the training and basic execution environment using PettingZoo \cite{terry_pettingzoo_2021}, a multi-agent reinforcement learning environment library based on OpenAI's popular Gym library. The PettingZoo environment is implemented as a parallel environment (rather than agent-environment cycle) in which all agents perform actions simultaneously before any observations are taken. We call our environment ``patrolling\_zoo'', and release it as an open-source project for others to use\footnote{\url{https://github.com/NU-IDEAS-Lab/patrolling_zoo}}.

Our training code is heavily based on that used by Yu et al. in their original MAPPO paper \cite{yu_surprising_2022}, though we have performed extensive work to integrate it with our ``patrolling\_zoo'' environment and modify as described in \cref{sec:training}, along with the implementation of new critic and GNN-based actor networks.


\section{Evaluation Methodology}
\label{sec:evaluation}

In this section, we describe our evaluation methodology and test setup. Importantly, while the training and initial evaluation of our approach was performed in a simple graph environment using the PettingZoo library, we also choose to evaluate our algorithm using a multi-agent robotics simulator which better highlights our approach's robustness to disturbances and applicability to real scenarios such as the multi-robot patrolling problem.

\subsection{Experimental Setup}
The primary tool used in evaluation is the Grex Machina multi-agent framework\footnote{\url{https://github.com/NU-IDEAS-Lab/grex}}, developed by the IDEAS Lab at Northwestern University for multi-agent research and based on the Robot Operating System~2 (ROS~2) \cite{macenski_robot_2022}.
Grex allows agents to be operated either in one of multiple available simulators (Gazebo, Flatland, etc.) or on physical robots, all without changing a single line of code.
The simulator that we select, Flatland, injects artificial Gaussian noise into sensor readings, creating a more realistic test of our algorithm than the PettingZoo training environment.


To test our algorithm on the patrolling scenario, we integrate existing state-of-the-art and benchmark patrolling algorithms and environments into the Grex framework. Many such algorithms and environments are from a patrolling simulator previously built by Portugal et al. \cite{portugal_distributed_2013}, which we have adapted and integrated into Grex\footnote{\url{https://github.com/NU-IDEAS-Lab/patrolling_sim}}. Portugal's original patrolling simulator was used in a large number of patrolling algorithms that are still considered to be state-of-the-art, including \cite{portugal_distributed_2013}\cite{portugal_cooperative_2016}\cite{farinelli_distributed_2017}\cite{wiandt_self-organized_2017}\cite{elgibreen_dynamic_2019}\cite{goeckner_attrition-aware_2024}.

Selection of appropriate algorithms for comparison with this work is critical, but is highly difficult due to varying assumptions and objectives between the previous approaches.
We compare with the AHPA algorithm \cite{goeckner_attrition-aware_2024}, SEBS \cite{portugal_distributed_2013}, and CBLS \cite{portugal_cooperative_2016}. All three benchmarks have strong performance in the face of agent attrition. However, none of them make use of environmental observations. To account for this, we first perform an experiment using unlimited observations and undisturbed communications, theorizing that this will show the best performance of all algorithms. We also attempt to find a middle ground with an observation radius of 40 m for subsequent tests and a variety of communication success rates. All algorithms are tested with and without attrition.

The primary metric that we use for performance comparison is the average idleness of all nodes $\hat{\zeta}$ over time. The use of this metric was established by \cite{portugal_distributed_2013} and it has appeared in almost all related works since that point. Therefore, we find it to be a suitable metric for judging the overall efficacy of our algorithm.

\subsection{Generalization}
\begin{wrapfigure}{R}{0.25\textwidth}
    \vspace{-10mm}
    \centering
    \includegraphics[width=0.20\textwidth]{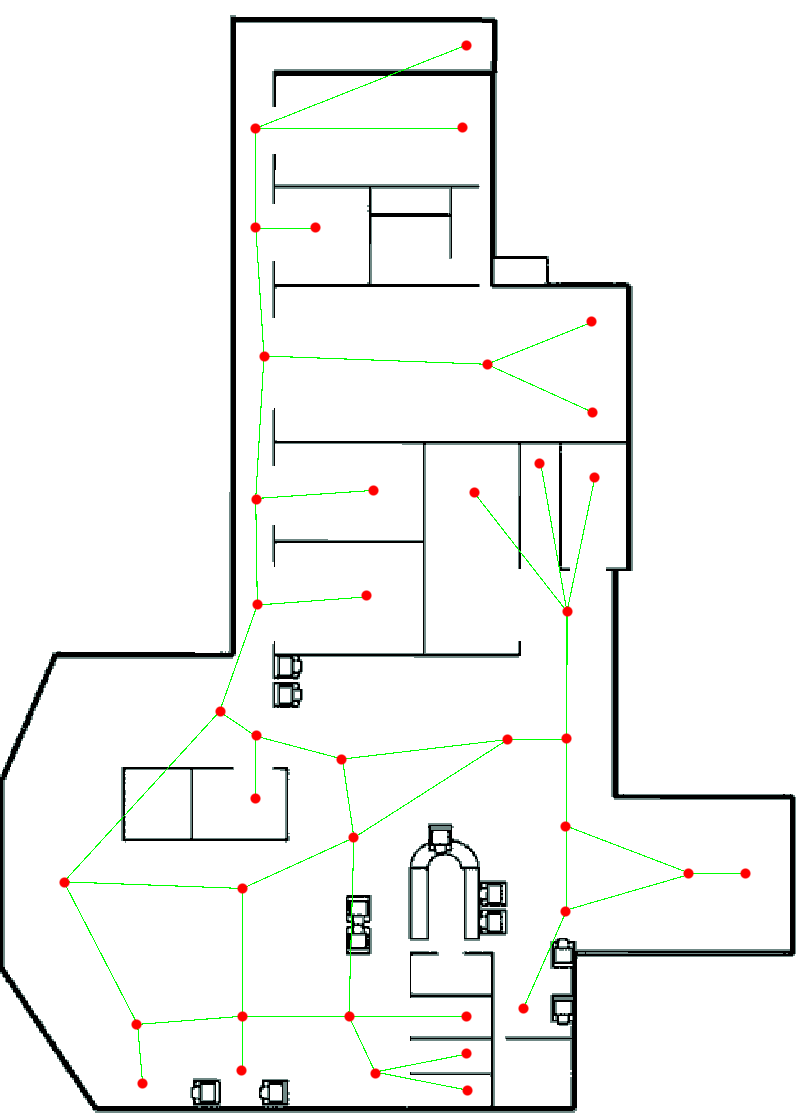}
    \caption{The agents were trained on the ``Milwaukee'' graph and successfully patrol on an entirely different one, ``Cumberland'' (above), demonstrating the generalizability of our approach. Above, red dots indicate nodes and green lines indicate edges in the patrol graph, while black lines indicate obstacles. Training is performed without obstacles, but in simulation, agents must avoid the walls.}
    \label{fig:cumberland}
    \vspace{-5mm}
\end{wrapfigure}

To ensure that the learned policy $\pi$ is not overfit and can generalize to different environments, we train and test with entirely different graphs and agent counts. For the results shown in this paper, training is performed on the ``Milwaukee'' graph with four agents. Evaluation is performed with six agents on the ``Cumberland'' graph (see \cref{fig:cumberland}), the same environment commonly used for testing in previous works \cite{portugal_distributed_2013}.

\subsection{Disturbances}
\label{sec:disturbances}
We model agent attrition in the simplest way possible: at two fixed points in time throughout the experiments, we choose an agent and remove it from the simulation. This simplicity provides good performance comparisons between algorithms by allowing the moment of attrition to be seen as an inflection point, where each algorithm either suffers in performance or continues unfazed.

As with attrition, we keep the communication disturbance as simple as possible, using a Bernoulli loss model with fixed reception probability. We ensure that all algorithms publish agent telemetry (positions, etc.) at a rate no more than one Hertz. Other messages, such as attrition notifications or goal-reached notifications, are sent on-demand. We apply the communication disturbance to all of these messages.


\section{Results}
\label{sec:results}

We find that our method is highly effective and suitable for use in multi-agent systems that must be fielded in environments where the risk of agent attrition or of communication disturbance is great. In this section, we present and analyze the findings of our work.

\subsection{Training Performance}

\begin{wrapfigure}{R}{0.30\textwidth}
    \centering
    \includegraphics[width=0.30\textwidth]{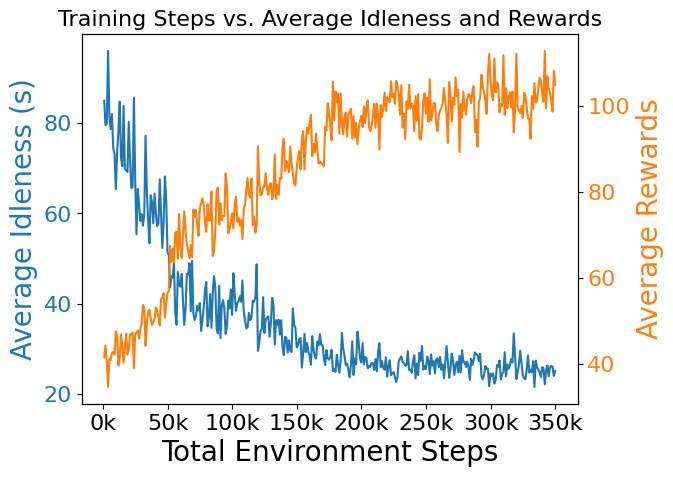}
    \caption{Graphs showing the average episode reward and evaluated average idleness over training period. Note that training completes in a mere total 350 thousand environment steps.}
    \label{fig:training_reward}
\end{wrapfigure}

The policy was trained for 350,000 environment steps, broken into environment episodes of 200 steps each, with five copies of the environment in parallel, for a total of 350 PPO episodes. In \cref{fig:training_reward}, both average reward and average idleness can be seen to improve and converge over time. Our improvements to MAPPO and the use of graph observations with a GNN greatly improved training time and feasibility over our previous (non-GNN) attempts. Experimentally, we found that using hyperparameters $\alpha=1.0$, $\beta=0.5$, and $\gamma=0.99$ resulted in the best training performance, effectively balancing global and local rewards.


\begin{figure}[h]
    \centering
    \includegraphics[width=0.45\textwidth]{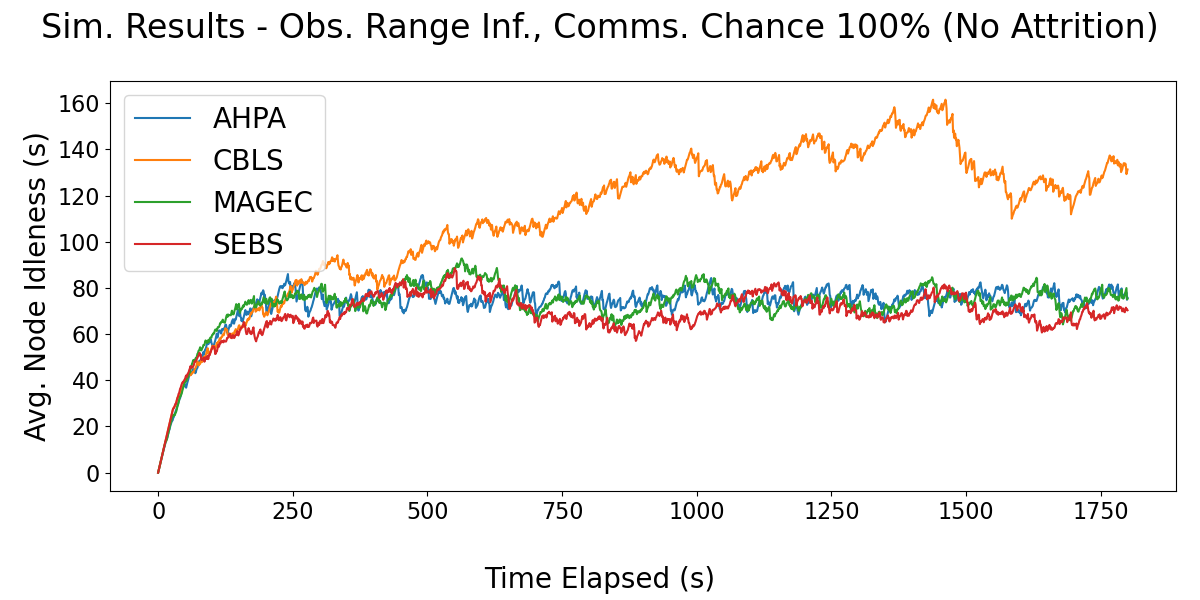}
    \includegraphics[width=0.45\textwidth]{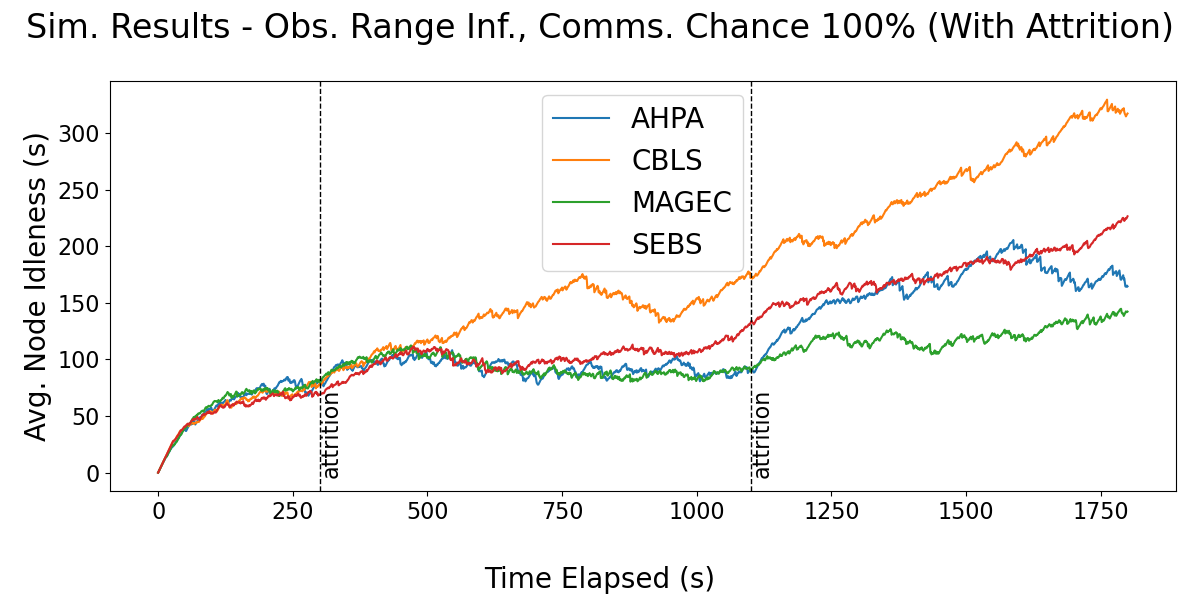}
    \caption{Test results using observation radius $\inf$ m and communication success rate of 100\%. At left, the test without attrition, and at right, the test with two attrition events. Our algorithm, MAGEC, is shown in green.}
    \label{fig:res_obsInf_comms100}
\end{figure}

\subsection{Simulation Performance}
In evaluation of our algorithm's performance in simulation, we focus on how the ``average idleness'' metric described in \cref{sec:patrolling} changes throughout the course of a 30-minute evaluation run. Due to artificial sensor noise in the simulator, we perform three evaluation runs of every experiment and then average the results across runs.

Though direct comparison between algorithms that make different assumptions is difficult, we believe that our method performs well. MAGEC achieves stable standard deviations of idleness throughout all tests, indicating that nodes in the graph are visited with similar frequencies. To provide the fairest comparison possible, we first test with an infinite observation range and $100\%$ communication success rate, on the theory that all algorithms will be able to achieve their maximum performance without being hampered by differing assumptions. As can be seen in \cref{fig:res_obsInf_comms100}, MAGEC performs far better than the existing algorithms in this scenario.

\begin{figure}[h]
    \centering
    \includegraphics[width=0.45\textwidth]{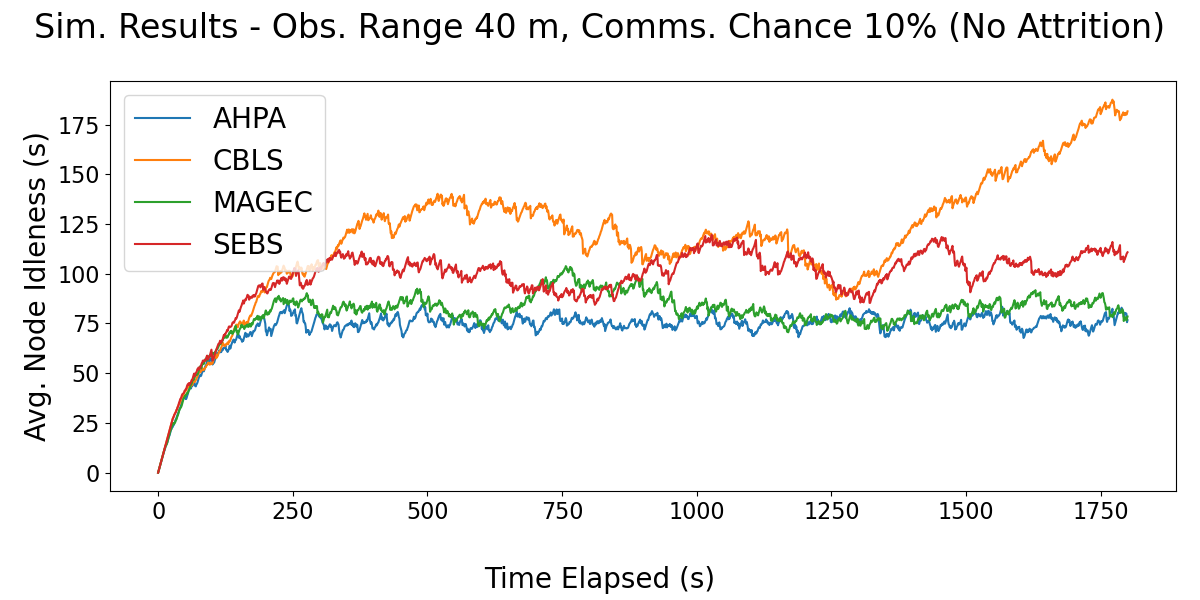}
    \includegraphics[width=0.45\textwidth]{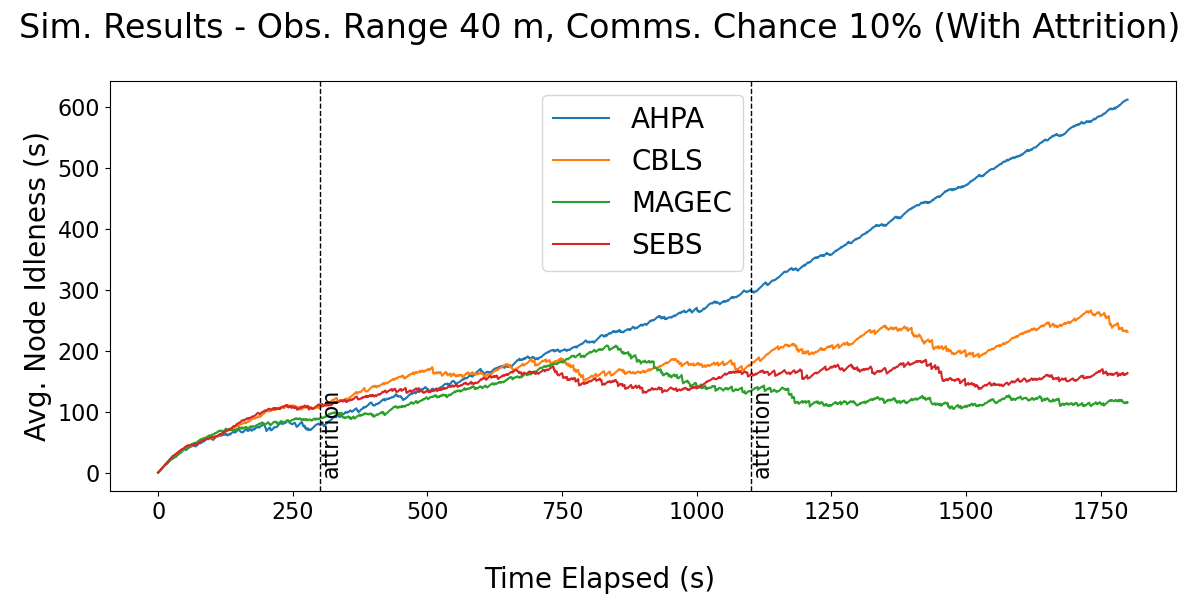}
    \caption{Test results using observation radius $40$ m and communication success rate of 10\%. At top, the test without attrition, and at bottom, the test with two attrition events. Our algorithm, MAGEC, is shown in green.}
    \label{fig:res_obs40_comms10}
\end{figure}

It also outperforms the benchmark algorithms in terms of average idleness, our primary metric, in attrition scenarios such as the one shown in \cref{fig:res_obs40_comms10} which uses a $10\%$ communication success rate and limited observation range.

Other algorithms do not appear to handle attrition well when faced with heavy communication losses. For example, AHPA handles attrition very poorly when its attrition notification message is lost, resulting in extremely high average idleness and standard deviation of idleness after attrition. CBLS also struggles when faced with message losses and attrition. However, SEBS fairs well and comes close to MAGEC in some scenarios.

\begin{figure}[h]
    \vspace{2mm}
    \centering
    \includegraphics[width=0.45\textwidth]{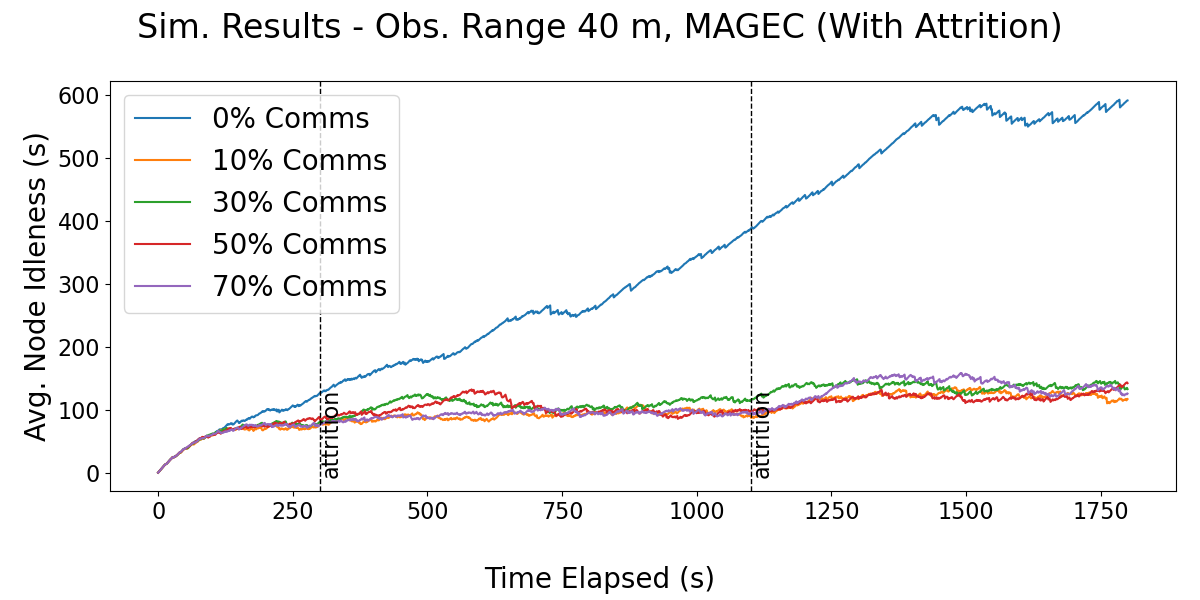}
    \caption{Test results using observation radius $40$ m and various communication success rates for MAGEC. Zero-communication performance is poor, but MAGEC's performance remains strong even with only 10\% comms.}
    \label{fig:res_commsTest}
    \vspace{-1mm}
\end{figure}

We also perform tests to determine MAGEC's performance with various observation ranges and communication success rates. As seen in \cref{fig:res_commsTest}, performance drops off significantly with zero communications, but otherwise is minimally impacted.

In non-attrition scenarios, MAGEC is competitive, even though it is sometimes outperformed by AHPA. We attribute this to AHPA's extremely deterministic patrol order, which results in a lower standard deviation of idleness and thus lower average idleness when attrition is not a factor. For AHPA, communication losses are also only important when attrition occurs, because AHPA does not use any messaging other than an attrition notification method. This makes AHPA a strong contender in the non-attrition, disturbed communications scenarios, but as expected, MAGEC performs best in the experiments involving agent attrition.

\section{Conclusion}

The performance of MAGEC, the Multi-Agent Graph Embedding-based Coordination algorithm, provides evidence that GNN-based MARL can be highly effective in an entire class of graph-environment problems such as multi-robot patrolling, vehicle routing, and swarm navigation. MAGEC is shown to effectively coordinate robots even with agent attrition, partial observability, and significant communication loss. Further, MAGEC operates in an environment with delayed rewards and sparse actions. These are realistic disturbances and limitations that multi-robot systems must overcome before they can be widely fielded. MAGEC provides a solid foundation for future work in this critical direction.

However, MAGEC is not magic, and more problems remain to be solved.
Future work may attempt to integrate better methods for prediction of unobservable state (such as that performed by SEBS) which will enable far better performance in limited-communication scenarios.

Regardless, our method is pioneering in its use of $k$-layer GNNs paired with MARL for multi-robot coordination. It represents a general solution for agents which must coordinate movement in any environment representable as a graph, and we hope that MAGEC will form the basis of many robust multi-robot systems to come.

\section*{Acknowledgment}
Special thanks to Mr. Yixuan Wang, Mr. Simon Zhan, and Dr. Stephen Xia for their advice on a million issues.


\bibliographystyle{ieeetr} 
\bibliography{IEEEabrv,refs}


\end{document}